\documentclass[11pt,twoside]{article}
\usepackage{newpasp}

\index{Borne, K. D.}
\index{Bushouse, H.}
\index{Colina, L.}
\index{Lucas, R. A.}

\begin{document}

\title{The Ultra-Luminous IR Galaxy Population}
\author{Borne, K. D.}
\affil{Raytheon ITSS and NASA Goddard Space Flight Center, Greenbelt, MD}
\author{Bushouse, H., Lucas, R. A.}
\affil{STScI, Baltimore, MD}
\author{Colina, L.}
\affil{Instituto de F\'{\i}sica de Cantabria (CSIC-UC), Santander, Spain}
\author{Arribas, S.}
\affil{Instituto de Astrof\'{\i}sica de Canarias, Tenerife, Spain}


\begin{abstract}
We briefly summarize some results from an on-going Hubble Space Telescope 
(HST) survey of a large sample of ULIRGs (Ultra-Luminous IR Galaxies).  New
ground-based multi-fiber spectroscopic observations are now being 
obtained to complement the HST data and to assist in the interpretation 
of these complex objects.
\end{abstract}




\section{Starbursting Galaxies and the ULIRG Population}

It has been known since the pioneering work of Larson \& Tinsley (1978)
that collisionally disturbed galaxies (e.g., Arp's [1966] peculiar
galaxies) have abnormally high star formation rates compared to
isolated galaxies of similar type.  It was not until IRAS discovered
the population of luminous IR galaxies (LIRGs) and ultraluminous IR
galaxies (ULIRGs), having extremely high star formation rates
(100--1000$\times$ the Galactic star formation rate), that the links
between strong star formation, high IR luminosity, and galaxy-galaxy
encounters were made inseparable (Joseph \& Wright 1985).  The
observation that some galaxies have higher rates of star formation than
can be sustained by their current gas content over a full Hubble time
led to the idea of "starbursting" galaxies (i.e., starbursts; 
Weedman et al.~1981).  
It has thus been unequivocally established that the
LIRG, ULIRG, starburst, and collision+merger processes are physically
related phenomena that are intimately connected to the star formation
history of galaxies, galaxy formation, and galaxy evolution (for a
review, see Sanders \& Mirabel 1996).  In fact, most recently, it appears
very likely that the cosmic IR background is produced by cosmologically
distant LIRGs and ULIRGs (i.e., dusty starbursts; Smail et al.~1998;
Blain et al.~1999; Barger et al.~1999).

\section{Results from a Survey of ULIRGs}

We have been studying a large sample of ULIRGs with both HST imaging
and ground-based spectroscopy (Borne et al.~1997a,b, 1998, 1999a,b,
2000; Colina et al.~1999, 2000; Arribas et al.~2000).  Among the
plethora of results being derived from this rich survey database, we
have found strong evidence for a multiple-merger origin for many of the
ULIRGs in our sample (Borne et al.~2000).  We have established a
morphological classification scheme for ULIRGs that indicates that the
sample is nearly equally divided between single objects (e.g., merged;
disturbed; IR-luminous QSOs) and multiple objects (e.g., pairs; compact
groupings; strongly interacting multiples).  We find very
little luminosity variation across these morphological
classes.  We have also verified the long-known belief that the ULIRG
population has a high interaction rate.  From a sample of nearly 130
ULIRGs, we find $\sim$98\% show evidence for close neighbors, tidal
disturbances, or on-going merging.  In
several cases, the galaxies were previously classified as isolated
and/or undisturbed from low-resolution ground-based imaging.  Our
sample is the largest that has been used to derive this interaction
rate estimate.  Most recently, we have been obtaining multi-fiber
integral spectroscopy for several ULIRGs,
which indicate that luminous gas-rich knots are ubiquitous
among these galaxies.  In every case, we find multiple line emission
sources, frequently in regions detached from the cores of the galaxies
or from any other region that is luminous in continuum light.  These
line-emitting regions have spectral characteristics of H\thinspace II
regions, LINERS, or AGN.  In some cases, these line-emitting clouds
may be simply reflecting the emission from a dust-obscured nuclear
source.  In general, we find that the ULIRG population of galaxies has
a rich dynamical diversity, demonstrated both morphologically and
spectroscopically.  This points to a rich evolutionary history for
these objects, with strong connections between the history of dust and
gas evolution, star formation, metal enrichment, and galaxy evolution.
The consequences of these connections may also be applicable to the
interpretation of distant submm/FIR sources.


\acknowledgements

Support for this work was provided by NASA through
grant number GO-06346.01-95A from the Space
Telescope Science Institute, which is operated by
AURA, Inc., under NASA contract NAS5-26555.

\end{document}